\documentclass[12pt,preprint]{aastex}

\newcommand{\kms}{km s$^{-1}$}

\shorttitle{Impact of a filament eruption on high lying loops}
\shortauthors{Harra et al.}

\begin{document}


\title{The impact of a filament eruption on nearby high-lying cool loops}


\author{L. K. Harra, S. A. Matthews  and D.M. Long}
\affil{UCL-Mullard Space Science Laboratory, Holmbury St Mary, Dorking, Surrey, RH5 6NT, UK }
\email{l.harra@ucl.ac.uk}

\author{G. A. Doschek}
\affil{Space Science Division, Naval Research Laboratory, Washington, DC 20375, USA}

\and

\author{B. De Pontieu}
\affil{Lockheed Martin Solar and Astrophysics Laboratory, Org. A021S, Building 252, 3251 Hanover Street, Palo Alto, CA 94304, USA}



\begin{abstract}
The first spectroscopic observations of  cool Mg II loops above the solar limb observed by NASA's Interface Region Imaging Spectrograph ({\it IRIS}; \cite{IRIS}) are presented. During the observation period IRIS is pointed off-limb allowing the observation of high-lying loops, which reach over 70~Mm in height. Low-lying cool loops were observed by the {\it IRIS} slit jaw camera for the entire 4 hour observing window.  There is no evidence of a central reversal in the line profiles and the Mg II h/k ratio is approximately  2.  The Mg II spectral lines show evidence of complex dynamics in the loops with Doppler velocities reaching $\pm$ 40 km/s.  The complex motions seen indicate the presence of multiple threads in the loops and separate blobs. Towards the end of the observing period, a filament eruption occurs that forms the core of a coronal mass ejection. As the filament erupts, it impacts these high-lying loops,  temporarily impeding these complex flows, most likely due to compression. This causes the plasma motions in the loops become blue-shifted and then red-shifted. The plasma motions are seen before the loops themselves start to oscillate as they reach equilibrium following the impact. The ratio of the Mg h/k lines also increases following the impact of the filament. 

\end{abstract}





\section{Introduction}
Cool loops have been observed in non-flaring active regions and in post-flare loop systems for many years. Complex motions are often seen in loops when the plasma cools, a phenomenon that is sometimes described as 'coronal rain'. These have been observed for many decades e.g. \citep{kawaguchi}. More recently, \citet{karel1} analysed data from the {\it TRACE} spacecraft and found that coronal rain in active regions was observed approximately every two days. Recent high resolution observations with {\it Hinode}, NASA's Solar Dynamics Observatory ({\it SDO} \cite{aia}), and instruments such as CRISP on the Swedish Solar Telescope have shown the coronal rain to consist of small and dense chromospheric cores with falling speeds of tens of \kms \citep{antolin}.  In these higher resolution datasets the rain appears to be ubiquitous, and it has been suggested that the \cite{karel1} observations with {\it TRACE} may have been picking up time periods when the blobs occur close together and in large quantities, in what is sometimes called a 'shower'. A range of speeds have been observed for the rain,  reaching a peak of 120 \kms , but with an average speed of around 60 \kms. Coronal rain can be most easily seen at the limb, but has been observed on the disk by \cite{antolin1}. The small-scale features  of the rain are important to understand as their descent may follow the magnetic field lines, and thus can provide information about the magnetic field structure. The rain is usually observed in chromospheric lines such as H$\alpha$ and Ca II H, and in absorption in EUV spectral lines. However, sources that resemble coronal rain have even been observed in white light during solar flares \citep{juan}. One interpretation for coronal rain is that hot loops will rapidly cool down through thermal conduction and radiation until becoming thermally unstable. This leads to the formation of the blob-like condensations.

Spectral line profiles of Mg II have rarely been measured above the limb. Skylab data were analysed \citep{uri} in the quiet Sun and an active region, and it was found that above 8$\arcsec$ the central reversal seen in the Mg II lines on the disk disappears.  A quiescent prominence was also observed in Mg II by the OSO-8 spacecraft \citep{vial} and blue-shifts were found reaching around 14\kms.  

 In this paper we study cool loops observed by {\it IRIS} on the 26th October 2013. IRIS is pointing off-limb, and hence allows us to explore the behaviour of cool loops lying above 30~Mm in height. These are the first observations of Mg II spectra at these altitudes. During the time of this observation a filament eruption occurs that disrupts these high-lying loops. We analyse the cool loops using the slit jaw data from IRIS together with spectroscopic measurement of Mg II before the eruption, and discuss the changes that occur as the eruption impacts these pre-existing loops. This is the first time that the impact of a filament eruption has been observed spectroscopically in these high lying cool loops. 

\section{Observations}
\begin{figure}
\plottwo{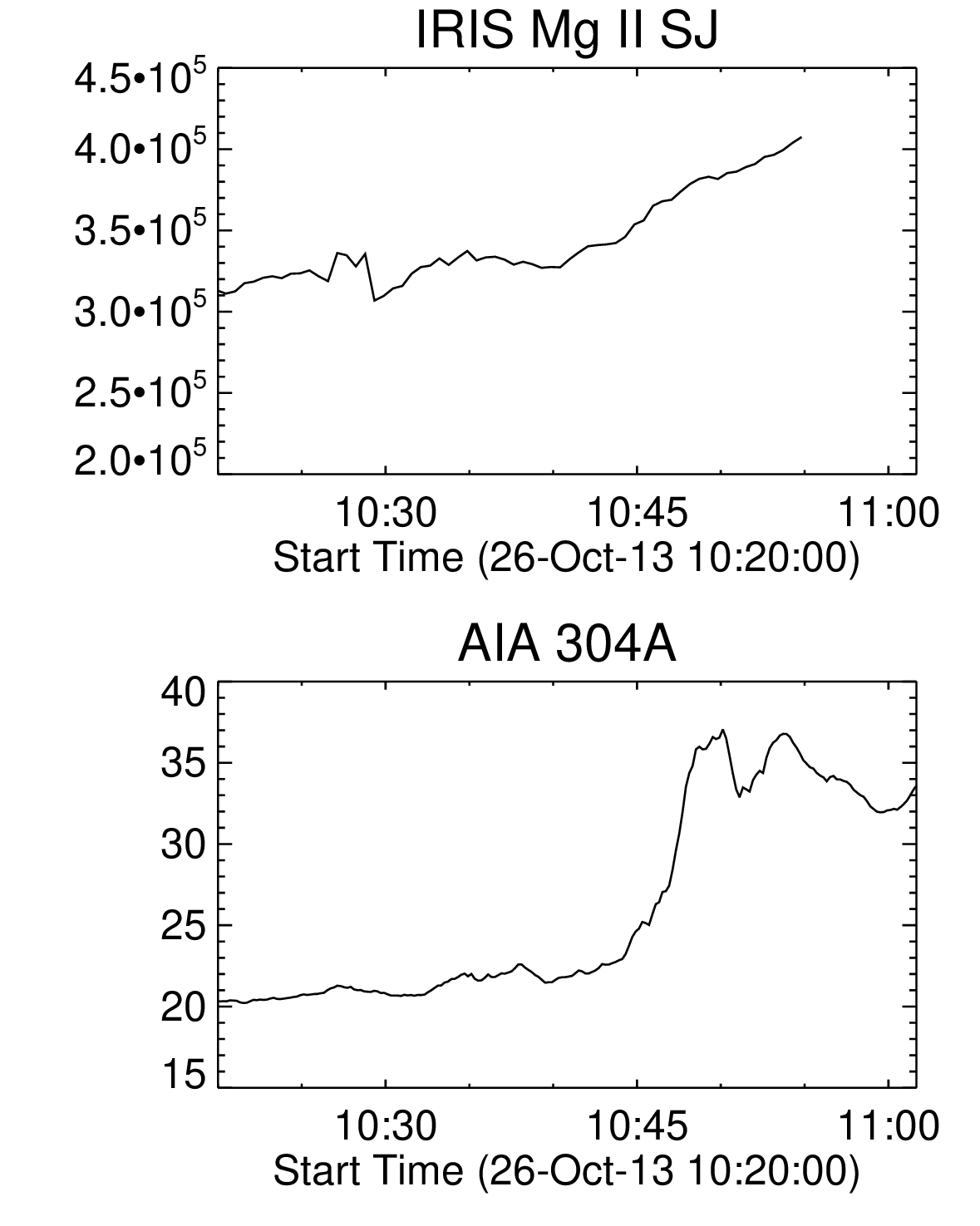}{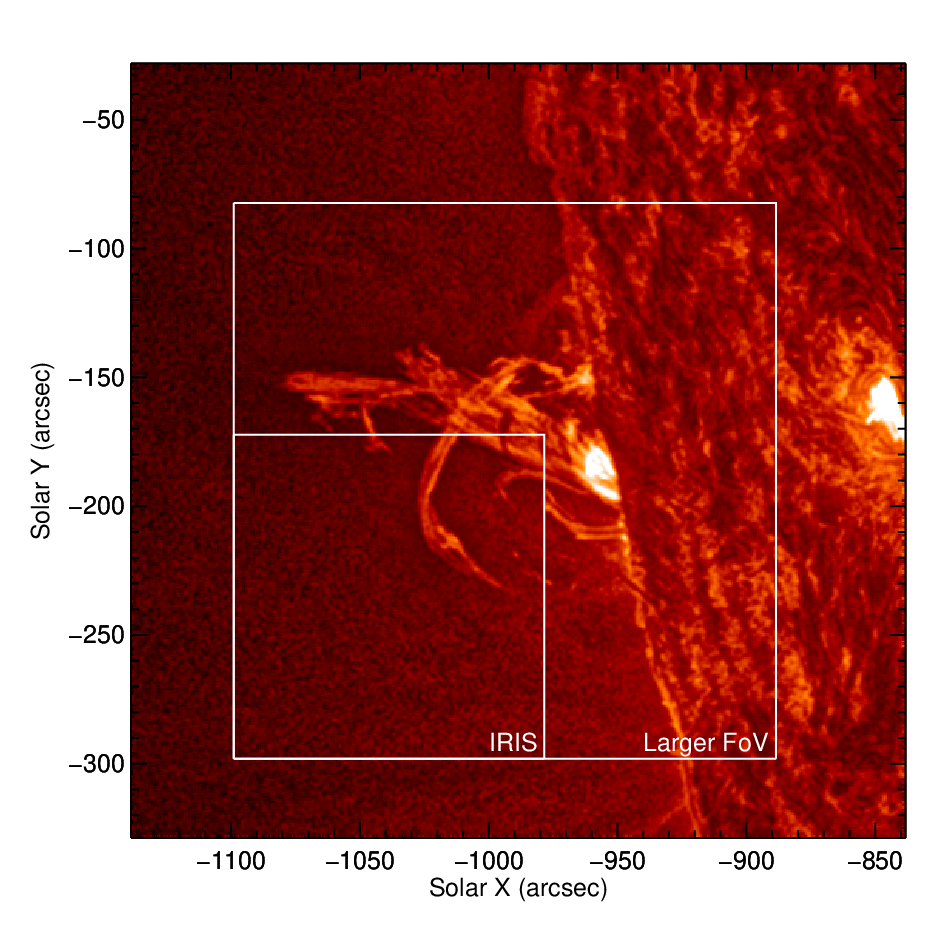}
\caption{The right hand image shows an {\it SDO} AIA image focussed on the active region at the limb. The left-hand side shows light curve from {\it AIA} 304 \AA\ passband from the larger FOV (lower plot) and light curve of the IRIS Mg II slit jaw data.   Movie 1 shows the AIA 304 \AA\ movie along with the IRIS Mg II movie. \label{aia_lcs}}
\end{figure}

We made use of datasets  from {\it SDO; \cite{aia}} and {\it IRIS}. {\it SDO} provides images of the full Sun in multiple passbands revealing the behaviour of plasma at different temperatures.  Figure~\ref{aia_lcs} shows the AIA 304 \AA\ light curve which indicates the activity levels of the filament as it erupts.  This figure also indicates when cool loops were lying at high enough altitudes to be seen in the IRIS field of view. {\it IRIS}  \citep{IRIS} provides  simultaneous spectra and images of the photosphere, chromosphere, transition region, and corona with 0.33--0.4 arcsec spatial resolution, two-second temporal resolution, and 1 \kms\ velocity resolution over a field-of-view of up to 175 arcsec $\times$175 arcsec. IRIS was launched into a Sun-synchronous orbit on 27 June 2013. The band passes include spectral lines formed in the chromosphere (Mg II h 2803  and Mg II k 2796 ) and transition region (C II 1334/1335  and Si IV 1394/1403 ). Slit-jaw images are taken simultaneously.  In our observations, the slit jaw camera observes from $\approx$ 35~Mm and the slit observed at $\approx$ 70~Mm above the limb. From Movie 1 it is clearly seen that there are cool loops during the whole observing time that lie below an altitude of 30~Mm. We analyse the time period from 10:20-10:54 UT that shows newly formed high lying loops, which are subsequently impacted by a filament eruption. Due to the high altitude of the observed loops, the count rates are very low in the C II, Si IV and Fe XII spectral lines, and so we focussed our analysis on the strong Mg II lines.  The analysis was carried out with the level 2 data-files as recommended in the {\it IRIS} data analysis guide. For the determination of the Doppler velocity we took an average of the spectrum along the slit. This gave us a rest wavelength of 2796.6 \AA. This is close to the National Institute of Standard and Technology wavelength in a vacuum for Mg II k of 2796.35 \AA.  The central reversal appears to have disappeared in the line profiles, consistent with earlier observations of this emission above the limb, indicating that the plasma is optically thin. This can be seen clearly in the spectra of the  'isolated' blob in Figure ~\ref{blob_stack}, which is described in the next section.  

\subsection{Characterising a single blob}

\begin{figure}
\begin{minipage}{\columnwidth}
\includegraphics[width=4.0in]{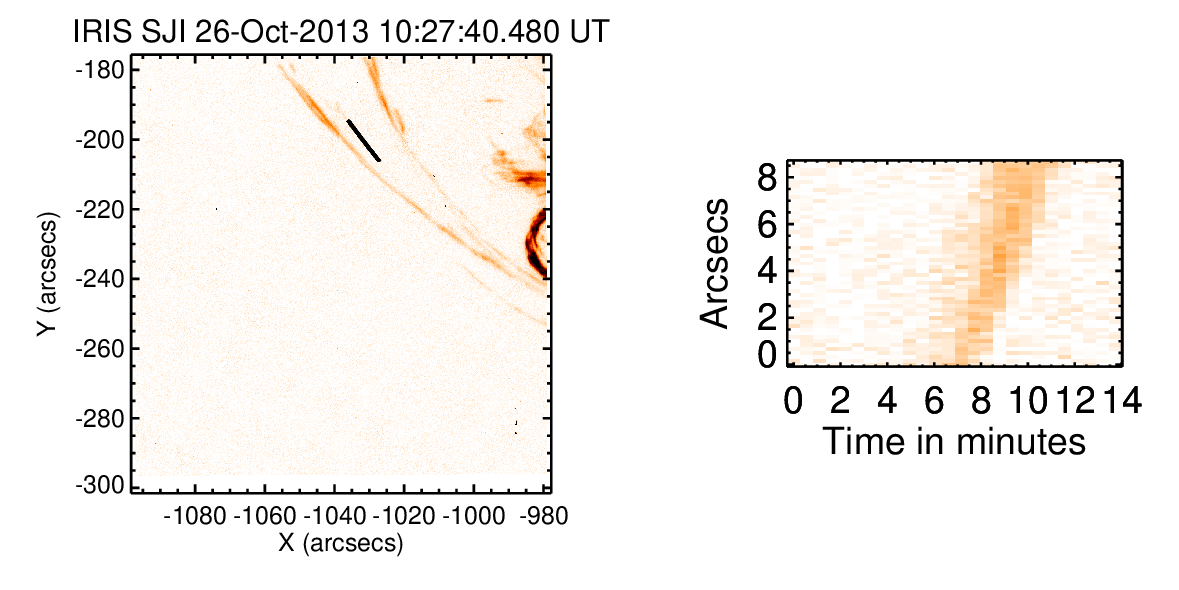}
\end{minipage}
\begin{minipage}{\columnwidth}
\includegraphics[width=4.0in]{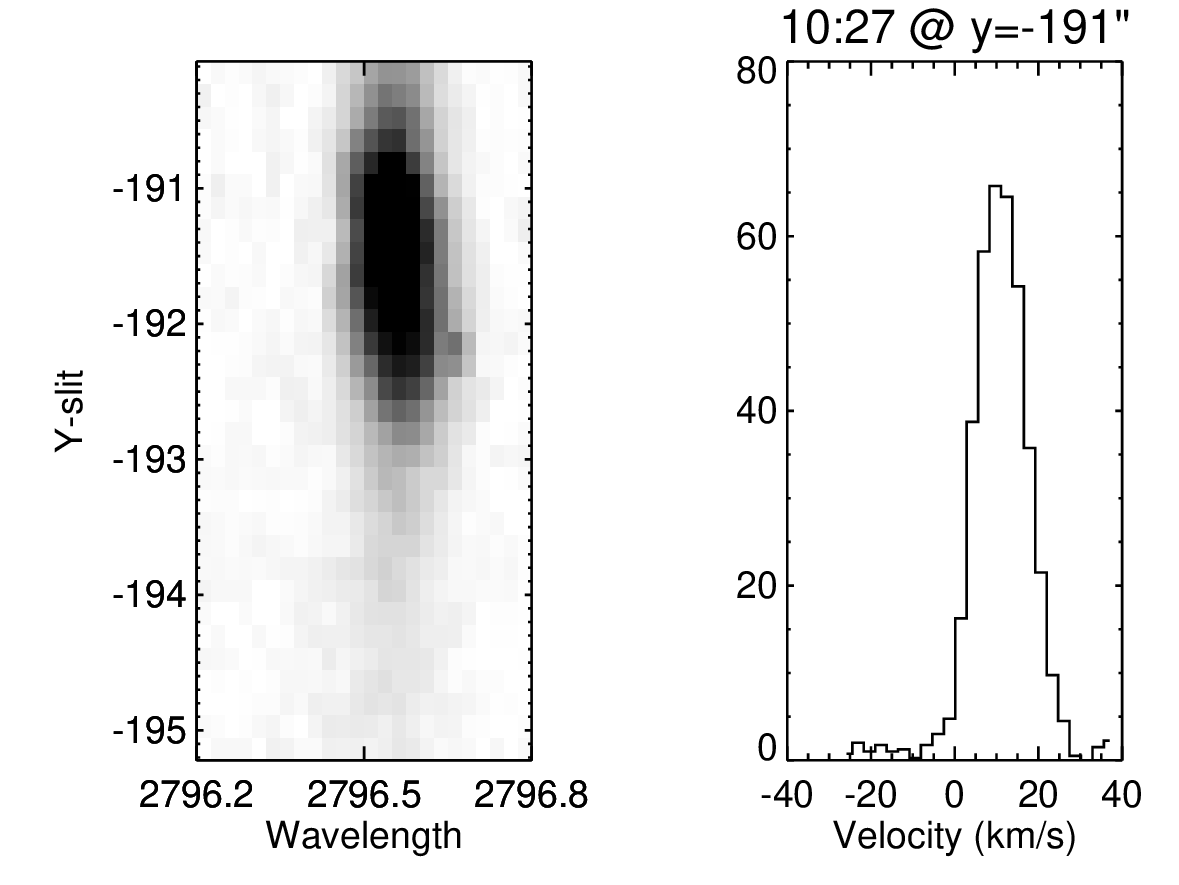}
\end{minipage}
\caption{The top left-hand figure shows the slit-jaw image in Mg II. The black line highlights a blob of falling, cool plasma that we studied. The top right-hand figure shows the motion of the blob (the zero arc seconds is the top of the black line). There is a clear propagation along this line with a speed of around 34 \kms. The bottom left-hand figure shows the spectra of Mg II along the slit as the blob falls downwards. The bottom right-hand figure shows sample spectra at 10:27 at the y-position 191 arc seconds. \label{blob_stack}}

\end{figure}

\begin{figure}
\epsscale{.80}
\plotone{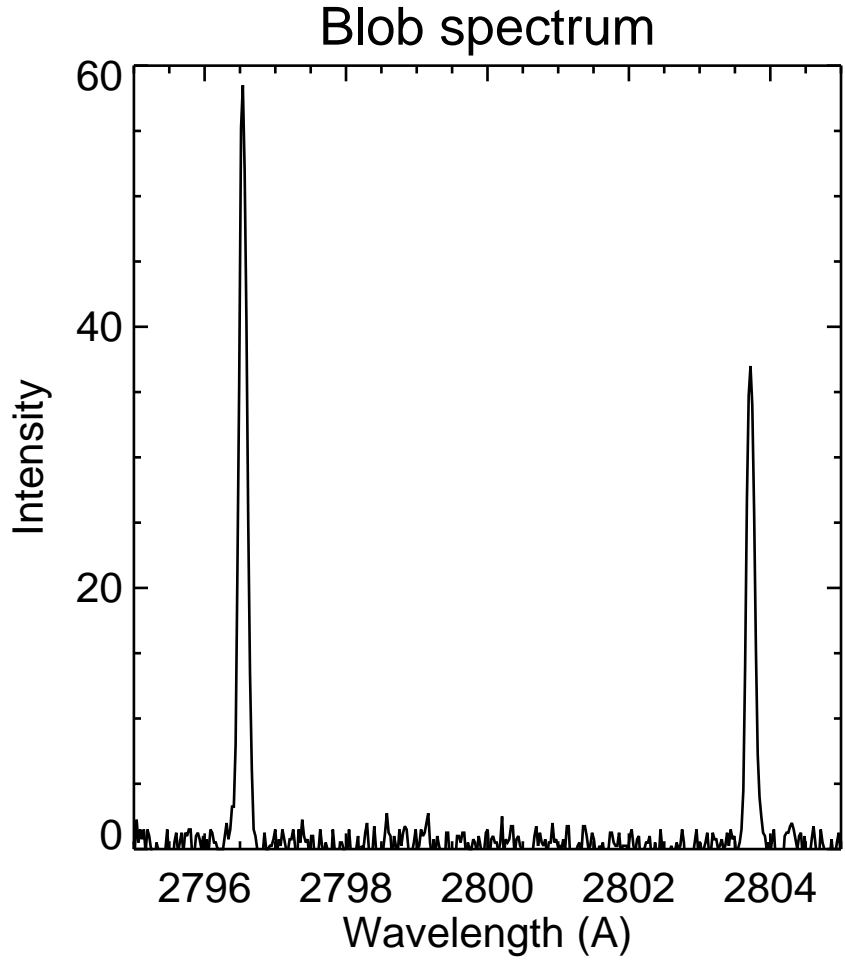}
\caption{A sample spectrum of the Mg II h and k lines in the isolated blob. The line intensity ratio of the two lines is 1.55. There is no evidence of central reversal in these lines.   \label{blob_ratio}}
\end{figure}

The structure of the cool loops is very complex -  there are multiple strands with a 'loop' structure, while at other times isolated blobs appear to follow the track of a magnetic loop. To explore the behaviour of the structures, we first isolate a single cool blob as it falls. Figure~\ref{blob_stack} shows an isolated blob of cool plasma falling downwards highlighted with the black line. The blob was tracked and  plasma flows were measured from the slit-jaw images.  There is a clear propagation along the black line, and the speed of the blob falling in the plane of the sky was measured to be $\approx$ 34 \kms. Figure~\ref{blob_stack} also shows the Mg II spectrum at the time the blob appears crosses the {\it IRIS} slit. The plot on the left shows this motion as a function of y-slit position and wavelength, while the right plot shows a sample spectrum. The spectrum is a simple Gaussian, with no evidence of multiple components and no evidence of central reversal of the lines. It is red-shifted with a speed of  $\approx$ 15 \kms. If we look at both the h and k lines, the ratio of the integrated spectral line intensities is 1.55. The ratio of the oscillator strengths of the h and k lines for simple electron impact excitation should be 2:1 - and for radiation then it should be 4:1. In this example the ratio is less than 2 (Figure~\ref{blob_ratio}), which suggest that radiative effects should be minimal, but may indicate the presence of resonance scattering.  Values such as this has been observed before, for example with Skylab data by \cite{george}, where values have been found which are 1.5 at 2\arcsec\ above the solar limb. More recent work by \cite{keenan} has shown for the first time that in the solar case, that this ratio can change in both directions as the result of opacity. 

This example was a rare isolated blob of short duration. The longer lasting loop structures show much more complexity in the brightness structures and flows,  which suggests that they may be composed of multiple component loops and falling blobs that are lying at slightly different angles to each other - hence the cool plasma falls at different orientations, each providing a different component to the Mg II profiles when the loops are so close together. This is suggestive of braiding which was predicted theoretically by \cite{eugene}. Braiding has been seen both in the chromosphere \citep{sarah} and more recently in the corona \citep{hic}. We discuss the temporal evolution of the more complex structure in the next section, and describe how the filament eruption changes the plasma behaviour in the loops.

\subsection{Temporal evolution of the plasma in the high-lying loops before and during the filament impact}

\begin{figure}
\begin{minipage}{\columnwidth}
\includegraphics[width=4.0in]{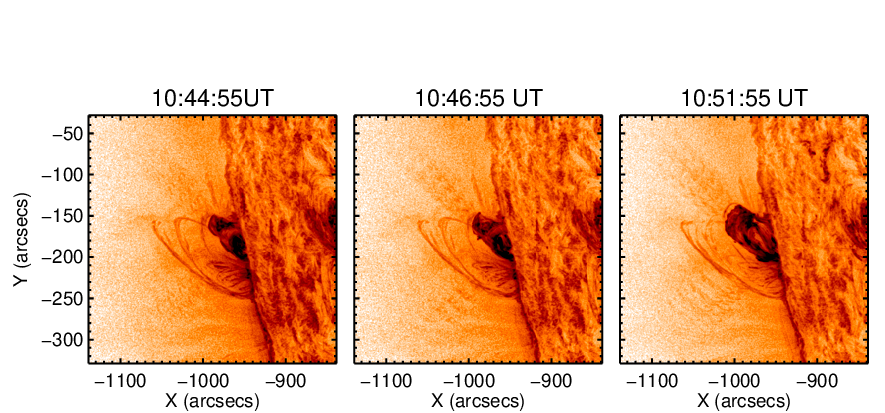}
\end{minipage}
\begin{minipage}{\columnwidth}
\includegraphics[width=4.0in]{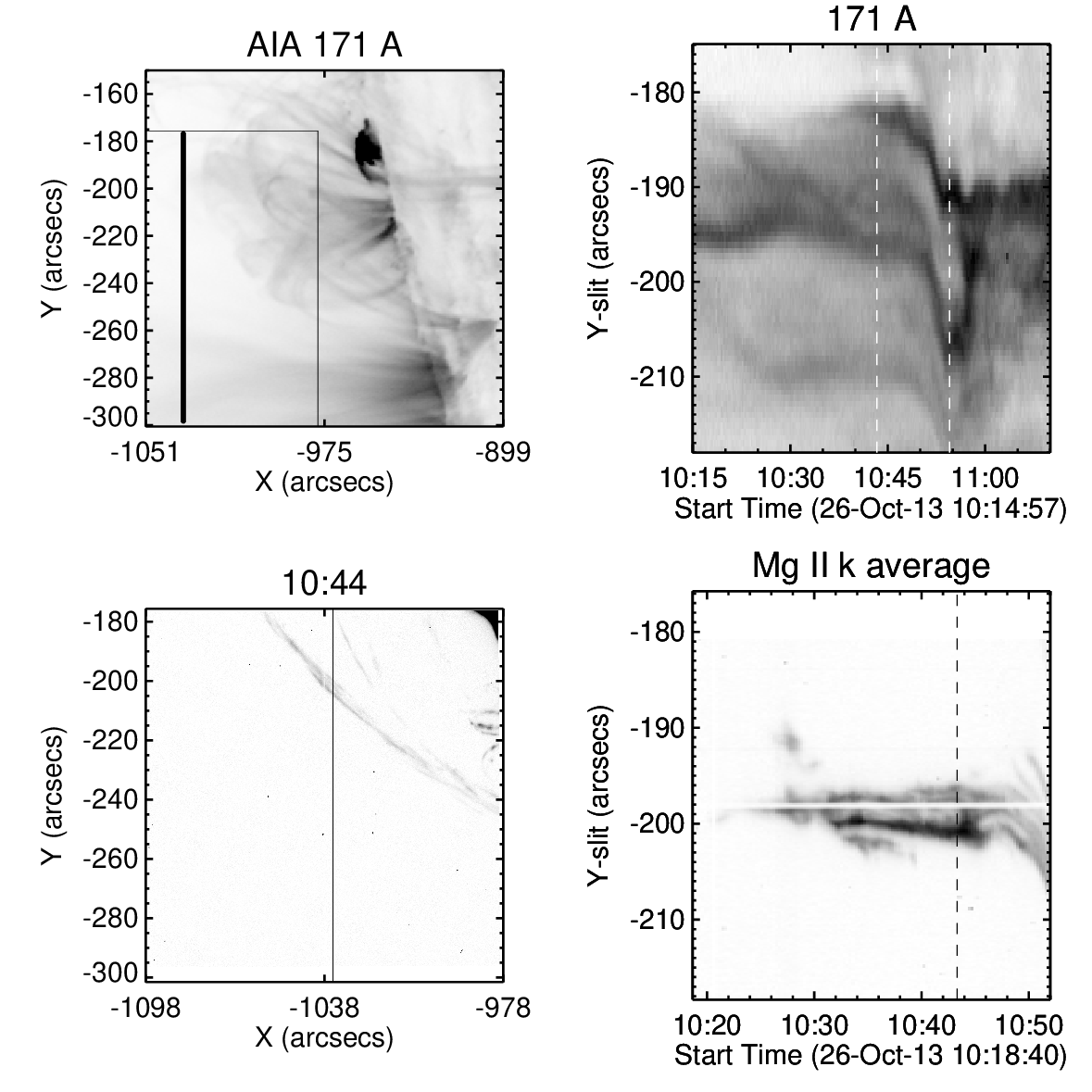}
\end{minipage}
\caption{The top three images show AIA 304 \AA\ data showing the filament eruption starting. The middle left image shows a 171 \AA\ AIA image with the {\it IRIS} field of view highlighted by a black box. The thick vertical black line shows the position where the AIA stack plot was made. The middle right hand plot shows the AIA stack plot. The first white vertical lines highlight the start of the eruption  and the second vertical lines indicates the end of the IRIS observations. {\bf Please note the y-axis has been changed to focus in on the dynamic features}.  The bottom lefthand image shows the {\it IRIS} slitjaw image at 2796 \AA\ showing the loops lying high in the corona.  The black vertical line shows where the slit is located. The bottom right image shows the {\it IRIS} slit data with time. The spectra have been integrated over wavelength to yield intensities. Distinct and broad features exist that appears to consist of many loops. Initially the loops are seen lying at roughly the same location but moving dynamically (at around -200$\arcsec$). Just before 10:50 UT the eruption propagates through and pushes the loops downwards by nearly 10 $\arcsec$ with a speed of 30 \kms. \label{sj_stack}}
\end{figure}

 Figure~\ref{sj_stack} shows example images from the {\it IRIS} slit jaw camera of the 10--20 x10$^3$ K Mg II and coronal AIA 171 \AA\ plasma. These images also show the location of the slit during this observation. The right hand panels of Figure~\ref{sj_stack} show a stack plot image of the Mg II k slit data with time where the slit crosses the cool loops. The loops seen are extremely dynamic showing fine structures and blob-like features falling continuously towards the solar disk as the loops cool (see Movie 1). The loops appear to  provide pathways for the cooling plasma to rain down into the lower solar atmosphere. This complex structure stays at the same location for a long time (more than 10 minutes) centred at -199$\arcsec$.  At around 10:48 UT the loops are moved in a southwards direction, which can be seen in the lower right stack plot that shows a downwards turning with speeds around 30 \kms. Figure~\ref{sj_stack} also shows the AIA 171 \AA\ data with a simulated stack plot in the same location. The IRIS observations of this region unfortunately stop at 10:54 UT before the loops have reached an equilibrium following the disruption. However the AIA 171 \AA\ data clearly shows how the loops recover with some oscillatory motion afterwards. These loop oscillations are known to be triggered by a nearby flare or eruption e.g. \citep{valery}. 

\begin{figure}
\epsscale{.95}
\plotone{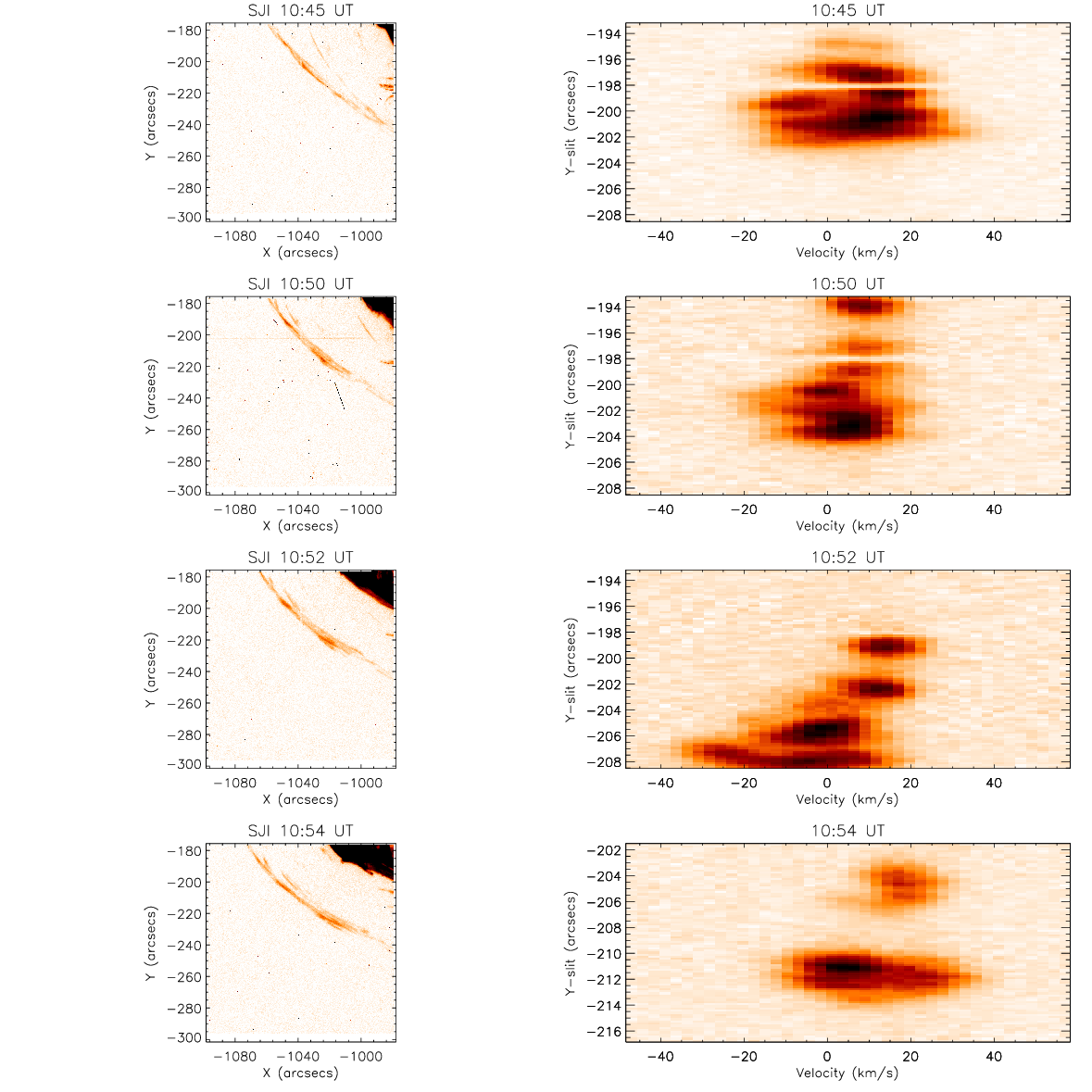}
\caption{In the left hand column the {\it IRIS} slitjaw images in 2796 \AA\ are shown at 10:45 UT, then at 10:50UT (when the eruption begins to push the loops), at 10:52UT and at 10: 54 UT. In the images the cool material that forms the core of the eruption is seen at the top right of the image. The hotter front is not seen at these wavelengths. The right hand side shows sample spectra of Mg II along the slit at each time. The spectra shown on the top right are at the same times as the images. The plasma in the loops show multiple features. At 10:50, once the eruption pushes the loops, the spectra are less complex with narrower lines profiles, indicating simpler dynamics. At 10:52 the plasma shows a strong blue-shift and at 10:54 the plasma then becomes red-shifted (please note that due y-axis has changed in order to track the feature). These strong flows are seen whilst the loops are being pushed downwards.  Movie 2 shows the spectra changing with time following the eruption. \label{IRIS_spectra}}
\end{figure}

Figure~\ref{IRIS_spectra} shows {\it IRIS} data from the slitjaw camera on the left and Mg II k spectral profiles on the right at four different times. The top images show the newly formed high-lying loops, and the next three images show the time period when the filament eruption impacted the loops. Movie 2 shows an animation of how the spectra change with time during the whole observing period.  The Mg II slitjaw images show the fine structure of the loops above the limb. However, the velocity scale indicates that speeds of $\approx$ 40 \kms\ blue-shifted (towards Earth) and $\approx$ 40 \kms\ red-shifted (away from the Earth) are present at different times.  In the first image showing the 'quiescent' cool loops, the structure is complex, with multiple 'layers' of threads and there is evidence of more than one component in the spectral lines at times. The complexity of the loop structure, including the falling blobs, is reminiscent of the coronal rain observed by \cite{antolin} where red and blue-shifted velocities of the same magnitude ($\approx$ 40 \kms) were observed in the H$\alpha$ line.

The filament eruption is first seen in the {\it IRIS} field of view at 10:45UT in the top right hand part of the image in Figure~\ref{IRIS_spectra} as a high intensity feature (clearly seen in Movie 1). At 10:52UT this is more prominent.  The high-lying loops are impacted  before this cool filamentary material reaches it, indicating that we do not see the front of the eruption at these cool temperatures. The sample spectrum at 10:50 UT shows the initial change that happens following the eruption impacting these loops. The spectra at this stage have narrower and less complex profiles, which indicates that some of the dynamics have been inhibited. At 10:52, the profiles show a significant change, now becoming predominantly blue-shifted. At 10:54 this has changed again with the profiles showing red-shifts.  These significant changes in the plasma dynamics are all occurring when the loops are still moving due to the eruption.  Movie 2 shows the {\it IRIS} Mg II k profile with time along the slit. 

\begin{figure}
\epsscale{.80}
\plotone{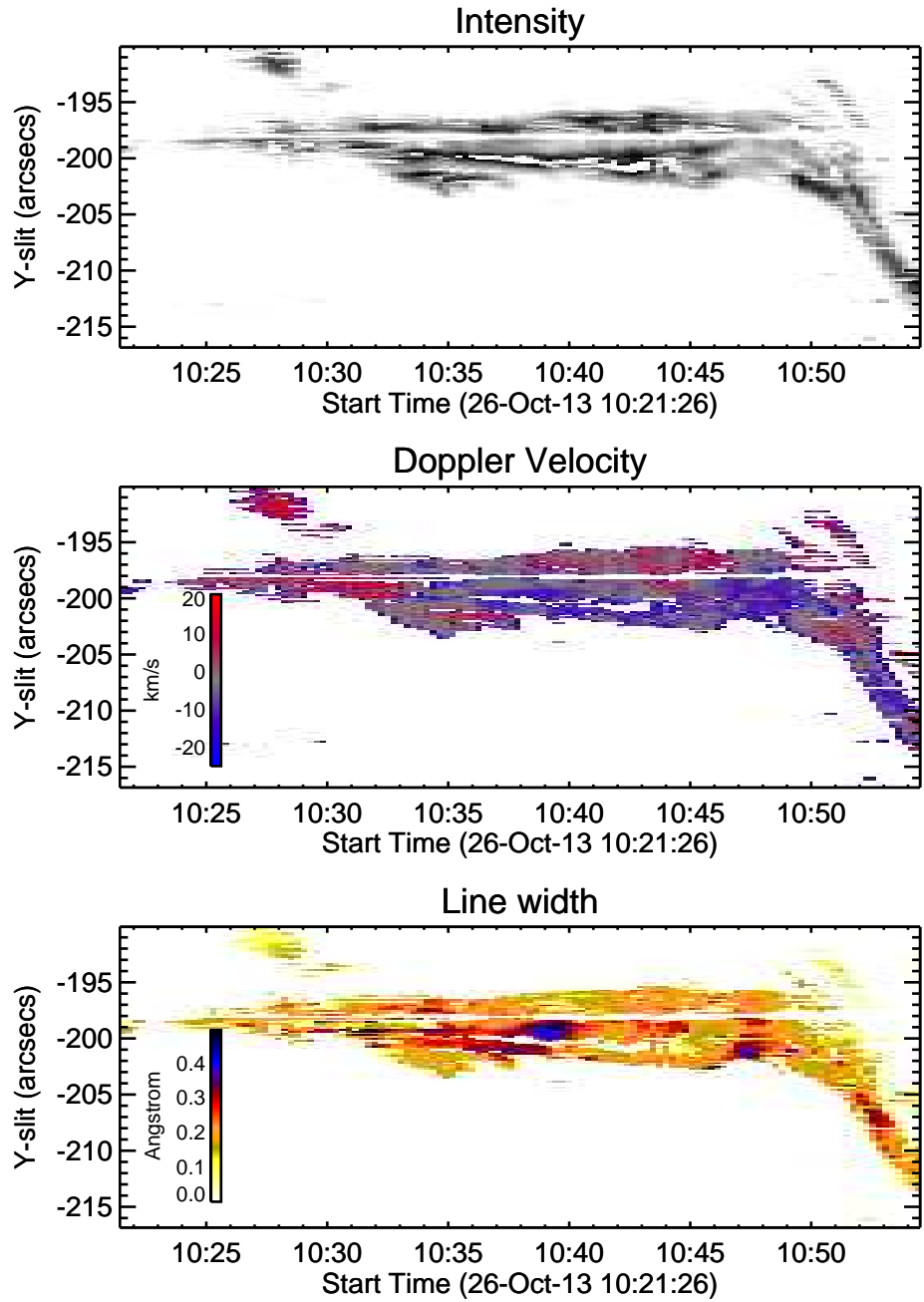}
\caption{A two component fit was applied for the MgII data. These figures show the Mg II slit data with time for the main fitted component - intensity is shown at the top, Doppler velocity in the middle and line width at the bottom. \label{2comp}}
\end{figure}

In order to appreciate the intricacy of the flows during this period, we fitted each spectrum with a two-component fit. The spectra are complex and several component fits were attempted, with the two component fit providing the best fit for the spectra. Figure~\ref{2comp} shows the intensity, Doppler velocity and line width results for the main component of the spectra. When these high loops initially form around 10:25UT, they are already showing flows (mostly red-shifts). As time progresses and more loops are created at these latitudes, there is a mix of red and blue-shifts. As the filament starts to impact the loops, there is an enhancement in line width (which indicates stronger flows in both directions) followed by a reduction in both line width and Doppler flows. The plasma then becomes, red-shifted, blue-shifted and then red-shifted again. The plasma is re-organising itself {\bf parallel} to the line-of sight before there is any evidence of the 'standard' loop displacements that have been observed following eruptions (\cite{valery}). This is the first time this has been observed. 

\begin{figure}
\epsscale{.80}
\plotone{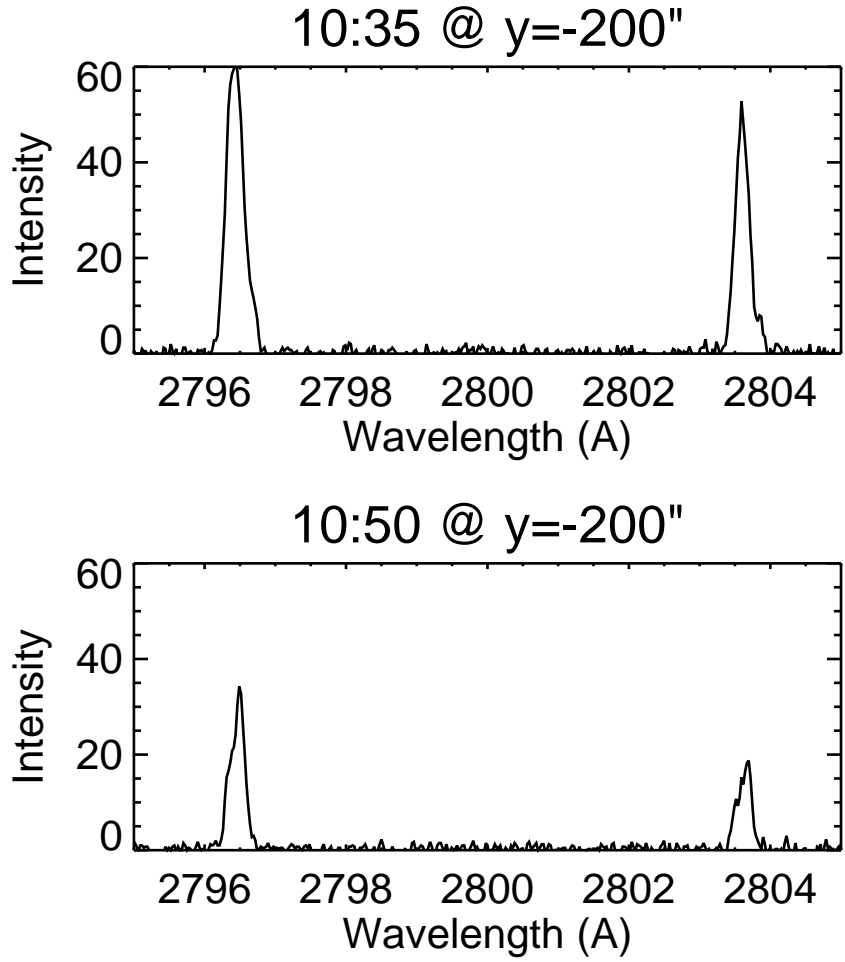}
\caption{Sample Mg II spectra at y=-200" before the filament eruption on the top, and after the filament eruption at the bottom.  \label{before_after}}
\end{figure}

\begin{figure}
\epsscale{.80}
\plotone{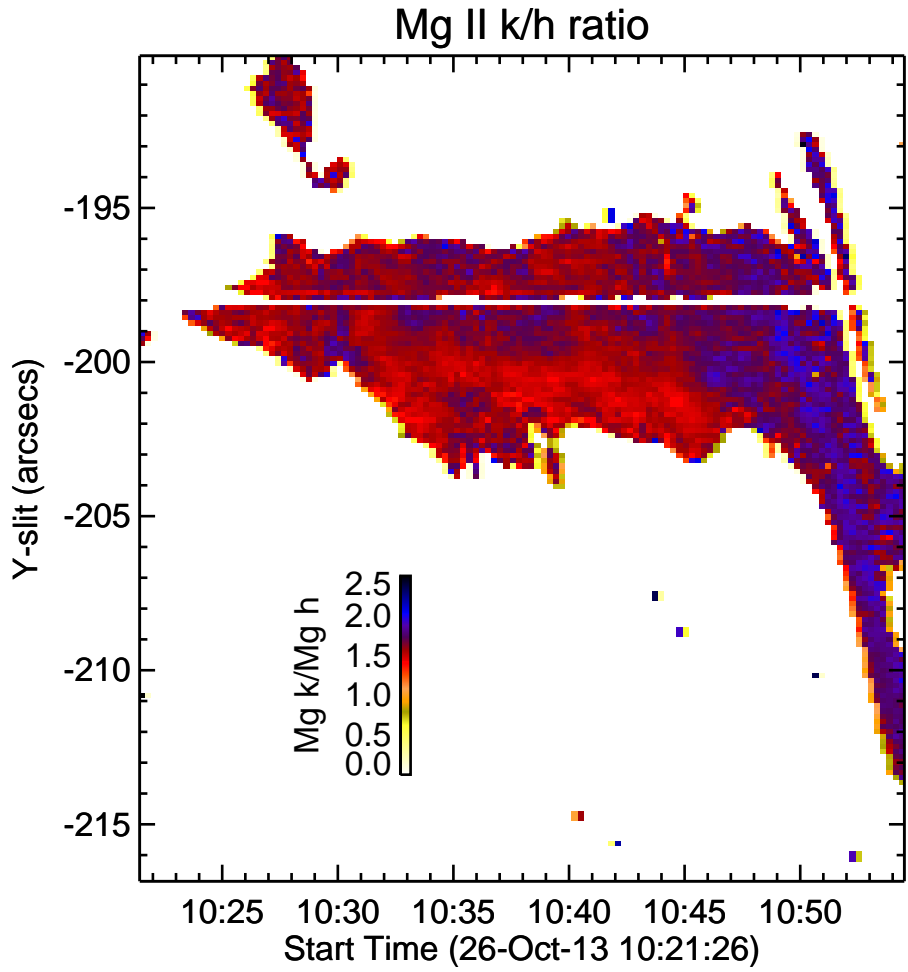}
\caption{A stack plot of the Mg k/h ratio. \label{intensity_ratio}}
\end{figure}

We determine the Mg II h/k ratio along the slit with time to determine if this changes during the impact of the filament. Examples of the Mg II h and k spectral lines are shown in Figure~\ref{before_after} spatially located in the centre of the loop structure. At 10:35 UT before the filament eruption, the profiles shown no central reversal. In addition the spectra are non-Gaussian broadened profiles compared with the spectrum of the isolated blob (Figure~\ref{blob_stack}). There is evidence of a red-wing component. The ratio of the h/k line at this time is 1.5 - similar to that of the isolated blob. At 10:50 UT just as the filament impacts these loops, the profiles again show complexity with stronger blue wing components. The ratio of this stage has increased to 1.9. To show the temporal and spatial evolution of the intensity ratio, we determine this value in all the pixels that have statistically significant intensity values. Figure~\ref{intensity_ratio} shows the stack plot of the ratio. Most of the pixels have a ratio around 1.6 before the filament impacts. The ratio increases to 2 and above following the impact. At no stage does the ratio reach values of 4 which would {\bf indicate} that radiation is the dominant process. 

We are seeing significant changes in these high-lying loops following the impact of the filament - both in terms of the complex dynamics within the loops and in terms of the line intensity ratios.

\section{Discussion}

These observations show coronal rain above the limb observed for the first time in Mg II by the {\it IRIS} spacecraft. The line profiles show that the central reversal is gone. The complexity seen in the spectral lines  can then be assumed to be due to dynamics only.
The spectral lines often show multiple components. We could in one instance isolate a blob of cool plasma falling, and found that its spectral profile was close to a single Gaussian, suggesting that in the other cases there were multiple threads overlaid in the same field of view, within the spatial resolution of {\it IRIS}. The ratio of the h/k lines is around 2 and shows variation with time and space. These loops then experienced the impact of a large eruption.   As the eruption propagates, it impacts the existing cool loops. The eruption has a cool filamentary core that is seen by {\it IRIS}, but the erupting plasma that initially reaches the cool loops is hotter and is not seen by {\it IRIS}.   As the eruption impacts these loops,  the complexity of the flows is temporarily reduced, most likely due to compression of the loops.  Then the plasma is red-shifted, blue-shifted and then red-shifted again. Alongside this, the line intensity ratio of h/k increases during the impact, reaching above 2:1.  It is possible that for this scenario, as the ratio becomes greater than 2:1 this might imply some additional radiative excitation along with the collisional excitation. The source of the radiative excitation above the limb may be enhanced by the flare and coronal mass ejection. Currently there are no simulations that make observational predictions of the response of these high-lying loops during the impact of the eruption, and these observations provide important constraints for future modelling work in this area. 

The {\it IRIS} data provides a microscope to the plasma dynamics of these cool high-lying loops during the filament eruption. From the images, we see clearly the loops disrupted by the filament eruption. In addition, {\it IRIS} demonstrates that the plasma inside reacts significantly to this event with the plasma reorganising itself before intensity oscillations occur. The level of turbulence going on inside loops during such a process has not been observed before, and allows us to probe plasma during a major disruption.


\acknowledgments
{\it Hinode} is a Japanese mission developed and launched by ISAS/JAXA, with NAOJ as domestic partner and NASA and STFC (UK) as international partners. It is operated by these agencies in co-operation with ESA and NSC (Norway). {\it IRIS} is a NASA small explorer mission developed and operated by LMSAL with mission operations executed at NASA Ames Research center and major contributions to downlink communications funded by the Norwegian Space Center (NSC, Norway) through an ESA PRODEX contract.
We thank the {\it IRIS} data for access to the data. The Solar Dynamics Observatory is a NASA mission. We thank the {\it SDO} team for access to the data.  The research leading to these results has received funding from the European Commission's Seventh Framework Programme under the grant agreement No. 284461 (eHEROES project).  GAD is supported by an Hinode grant from NASA. 





\clearpage



\clearpage









\clearpage

\end{document}